\documentclass[aps,twocolumn,showpacs]{revtex4} 
\usepackage{graphicx}

\newcommand{\sfig}[2]{
\centerline{ 
\includegraphics[width=#2]{#1}
		}
	}
\newcommand{\Sfig}[2]{
	\begin{figure}
	\sfig{#1}{0.95\columnwidth}
	\caption{{\small #2}}
	\label{fig:#1}
	\end{figure}
}
\newcommand{\Rf}[1]{\ref{fig:#1}}

\def\cmm2{{\,\rm cm^{-2}}}
\def\cm2{{\,{\rm cm}^2}}
\def\cmm3{{\,{\rm cm}^{-3}}}
\def\gcmm3{{\,{\rm g\,cm^{-3}}}}

\def\fun#1#2{\lower3.6pt\vbox{\baselineskip0pt\lineskip.9pt
  \ialign{$\mathsurround=0pt#1\hfil##\hfil$\crcr#2\crcr\sim\crcr}}}

\def\etal{{\it et al. }}

\def\be{\begin{equation}}
\def\ee{\end{equation}}
\def\bea{\begin{eqnarray}}
\def\eea{\end{eqnarray}}

\newcommand{\ec}[1]{Eq.~(\ref{eq:#1})}

\newcommand{\eql}[1]{\label{eq:#1}}

\newcommand{\omnu}{\Omega_\nu}
\newcommand{\omde}{\Omega_{\rm DE}}

\begin{document}

\title{Neutrino Mass and Dark Energy from Weak Lensing} 

\author{Kevork N. Abazajian$^1$ and Scott Dodelson$^{1,2}$}

\affiliation{{}$^1$NASA/Fermilab Astrophysics Center
Fermi National Accelerator Laboratory, Batavia, IL~~60510-0500\\
{}$^2$Department of Astronomy \& Astrophysics, The University of Chicago, 
Chicago, IL~~60637-1433}

\date{February 4, 2003}

\begin{abstract}
Weak gravitational lensing of background galaxies by intervening
matter directly probes the mass distribution in the universe. This
distribution, and its evolution at late times, is sensitive to both
the dark energy, a negative pressure energy density component, and
neutrino mass. We examine the potential of lensing experiments to
measure features of both simultaneously. Focusing on the radial
information contained in a future deep $4000$ square degree survey, we
find that the expected ($1$-$\sigma$) error on a neutrino mass is
$0.1$ eV, if the dark energy parameters are allowed to vary.  The
constraints on dark energy parameters are similarly restrictive, with
errors on $w$ of $0.09$. Much of the restrictive power on the
dark energy comes not from the evolution of the gravitational
potential but rather from how distances vary as a function of redshift
in different cosmologies.
\end{abstract}
\pacs{98.80.-k,98.65.-r,14.60.Pq  \hspace{4.6cm} FERMILAB-Pub-02/331-A}

\maketitle

{\parindent0pt\it Introduction.}  Rapid advances in astronomical observations
have enabled us to learn about the {\it dark sector} in the universe, both dark
matter that dominates over ordinary baryonic matter and dark energy which
apparently pervades the universe, with energy density a factor of two larger
than that of matter. Some of the dark matter is in the form of massive
neutrinos. We know that neutrinos have mass because we have seen evidence for
the transformation of one species into another\,\cite{fukuda,sno,lsnd}, a
transformation that is impossible for massless neutrinos. The standard
cosmology \,\cite{mc} predicts a definite relation between the cosmic neutrino
abundance and the cosmic photon abundance. Since the latter is well-measured, a
non-zero neutrino mass translates into an unambiguous prediction for the energy
density contributed by massive neutrinos.  The evidence for the existence of
dark energy is twofold.  First, distant Type Ia supernovae are fainter than
they would be if the universe were decelerating\,\cite{riess}, and acceleration
can take place only with negative-pressure dark energy. Second, observations of
anisotropies in the cosmic microwave background (CMB)\,\cite{cmbflat,wmap}
confirm that the universe is flat (so that the total density is equal to the
critical density), while many independent measurements\,\cite{matter} place the
matter density at one third of the critical density: the remaining two-thirds
is called dark energy.

There is still a great deal of uncertainty regarding both neutrino masses and
dark energy.  Both phenomena await a convincing theoretical interpretation in
the context of particle physics models. Beyond this theoretical uncertainty,
several relevant parameters have not been well measured. There is a wide range
of allowed neutrino masses: neutrinos might contribute as much as $20\%$ of the
matter density in the universe\,\cite{2df} or as little as $0.3\%$.  The
pressure of the dark energy is constrained to be negative, but how negative is
still unknown. In particular, a cosmological constant has $w=-1$ where $w$ is
the ratio of pressure to energy density. Many dynamical models however predict
much different values of $w$.  Better measurements of each of these fundamental
parameters, the energy density in massive neutrinos and the equation of state
of the dark energy, are clearly needed.

Remarkably both of these new pieces of physics\,---\,neutrino masses
and dark energy\,---\,leave similar signatures in the matter distribution
in the universe. In particular, how rapidly structures grow is determined
by the energy content of the universe. Since neutrinos are somewhat
relativistic even at late times, they inhibit the growth of structure.
Similarly, since the dark energy does not cluster, growth in the dark matter
slows in a dark-energy dominated universe. If one could measure
how rapidly the gravitational potential was evolving in time, one could
learn about this new physics.

Weak gravitational lensing offers the promise of making measurements
of precisely this evolution. In 2000, four groups\,\cite{weaklensing}
announced detection of the distortion of the shapes of distant
galaxies due to intervening large scale structure. Since then, the
observations have steadily improved, prompting a number of proposals
for larger and deeper surveys.  Here we explore the potential that
these surveys have for measuring neutrino masses and properties of the
dark energy\,\cite{hannestad,wlw}. Although the angular correlations of
the galaxy ellipticities contain useful cosmological information, we
focus on the radial information: the change in the shear field for
background galaxies in different redshift bins.  Radial tomography has
recently been applied to measure the properties of the shear field of
a cluster of galaxies~\cite{wittman}.

{\parindent0pt\it Lensing of Galaxies at Fixed Redshift.}  Hu
\cite{Hu1999,Hu2002} pointed out that, by breaking up the background
galaxies in a wide, deep survey into redshift bins, one could
essentially do tomography. The deeper bins probe an integral of the 3D
gravitational potential over distances farther away from us than the
nearer bins.

With many background galaxies at a fixed redshift $z_s$, one can hope
to recover the lensing convergence $\kappa$ in different angular
pixels.  The convergence in any one of these bins is sensitive to the
matter distribution between us and the sources at $z_s$. In
particular, 
\be 
\kappa(z_s,\vec\theta) = \int_0^{z_s} dz P(z,z_s)
\delta(z,\vec\theta)\,, \eql{kapdel}
\ee 
where $\delta$ is the fractional
deviation of the density from its average value and $P$ is the kernel
relating this deviation to the convergence. If we discretize
\ec{kapdel} so that $\kappa_i= \left[P_{\kappa\Delta}\right]_{ij}
\delta_j$ (where $i,j$ label different redshift bins), then the
projection operator is 
\be 
\left[P_{\kappa\Delta}\right]_{ij} =
\cases{ {3\over 2} \Omega_m H_0^2 \delta \chi_j {(\chi_{i+1}-\chi_j)
\chi_j\over \chi_{i+1}} & $\chi_{i+1} > \chi_j$\cr 0 & $\chi_{i+1} \le
\chi_j$} 
\ee 
where $\chi_i$ is the comoving distance out to redshift
$z_i$ and $\delta\chi_j$ is the comoving width of the $j$th redshift bin.
We take bins of width $\delta z=0.1$ and angular bin size of 1 deg$^2$ in
all that follows. Comoving distance depends on the energy density in the
universe at various times; it can be expressed as 
$\chi_i =\int_0^{z_i} {dz/H(z)}$, where $H$ is the expansion rate of the
universe.  The bottom panel of Figure~\Rf{lensing} shows the
projection operator for background galaxies at redshift two. The
convergence signal is most sensitive in this case to the matter
distribution at redshift $0.5$. As the background galaxies move closer
to us, they probe the structure at lower redshifts, later times.

It is possible to invert \ec{kapdel} and extract an estimate of
$\delta$ from the measurements of convergence\,\cite{keeton}.  On large scales,
$\delta$ is drawn from a Gaussian distribution with mean zero and a
variance which evolves with time. How this evolution takes place
depends upon the underlying cosmology; this dependence is expressed in
the {\it growth function}, $D(a=(1+z)^{-1})$, which is governed by\,\cite{mc}
\be 
D'' + \left(\frac{H'}{H}+\frac{3}{a}\right) D' - \frac{3}{2}
\frac{\Omega_m}{a^5} \frac{H_0^2}{H^2} D = 0
\eql{Mgrowth}\,,
\ee 
where $\Omega_m$ is the matter density today in units of the critical
density.  When neutrino masses are introduced, the growth function
becomes more complicated, and varies for modes with different
wavelengths. To a rough approximation, a nonzero neutrino mass
produces a fractional decrease in the power on scales probed by
lensing surveys equal to $12f_\nu$, where $f_\nu\equiv\omnu/\Omega_m$,
the ratio of the massive neutrino energy density to that in matter.
More accurate fitting formulae for the growth function and power
spectrum were calculated including nonzero neutrino mass in
Ref.~\cite{hueis}.  In the standard cosmology, $\omnu$ due to one
massive neutrino species is equal to $m_\nu/(94 h^2\,{\rm eV})$ where
$h$ parametrizes the Hubble constant. For a flat universe, then, 
\be 
f_\nu = 0.081 \, {m_\nu\over 1\,{\rm eV}}\,
{0.13\over \Omega_m h^2} .
\ee 
The combination $\Omega_mh^2$ is well-determined by CMB
experiments\,\cite{netterfield,pryke}; it is currently currently 
measured to be $0.13\pm0.01$\,\cite{spergel}.

The convergence then depends on the expansion history of the universe
via both the growth function and the projection operator.  The
Friedman equation expresses the expansion rate in terms of energy
density
\be
{H(z)\over H_0} = \left[ (1-\omde) (1+z)^3 + \omde (1+z)^{3(1+w)} \right]^{1/2}\,,
\ee
where $\omde$ is the dark energy density, and $w$ is the dark energy
equation of state.  In principle, we can hope to measure $\omde,w,$
and $f_\nu$ from a deep, wide lensing survey.

\Sfig{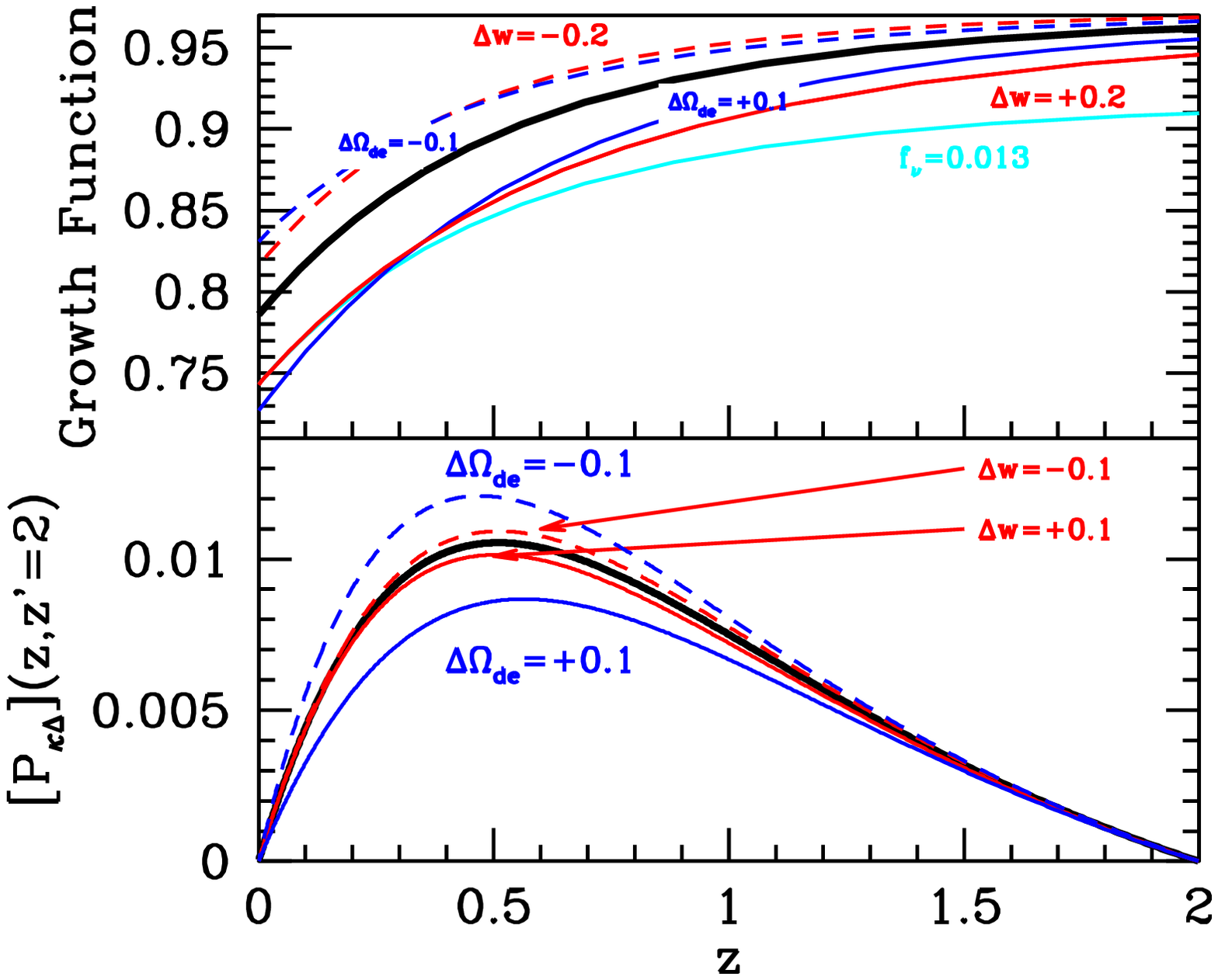}{{\it Top panel.} Growth function vs. redshift for
different choices of cosmological parameters. The base model in both
panels (dark solid curve) has $\omde=0.65, w=-1$ and
$f_\nu=0.005$. Negative changes in the parameters are indicated by
dashed curves. {\it Bottom panel.} Projection kernel as a function of
redshift for background galaxies at redshift $z'=2$. 
The projection is completely independent of neutrino
mass. Weak lensing convergence is the convolution of the two panels.}

The top panel in Figure~\Rf{lensing} shows the growth function for the base
model we use throughout, one in which $\omde=0.65,w=-1,\Omega_m h^2=0.13,$ and
$f_\nu=0.005$. Also shown are the slight changes in the growth function as
these parameters vary. Note that the changes induced by $\omde$ and $w$ are
very similar to each other. The bottom panel illustrates that the projection
operator depends almost exclusively on the dark energy density.

{\parindent0pt\it Results.}  As mentioned above, there are two effects
measured in a tomographic weak lensing survey. The first is the
evolution of the power spectrum, or the growth function, while the
second is the projection of physical distances onto redshift
space. Figure~\Rf{projection} shows the relative constraining power of
these two effects in a $4000$ square degree survey which measures
ellipiticites and (photometric) redshifts for a hundred galaxies per
square arcminute.  We use a galaxy redshift distribution of $dN/dz
\propto d\chi/dz \exp[-(\chi/\chi_\ast)^4]$, such that $\chi_\ast$
gives a mean redshift of one, but the results are insensitive to this
choice\,\cite{depth}.  We assume a constant effective equation of
state $w$, and an intrinsic r.m.s. galaxy ellipticity of 0.3.  The
inner region in Figure~\Rf{projection} includes both effects; the
outer ignores projection effects. The growth function depends on only
a combination of $\omde$ and $w$: a change in one can be offset by a
change in the other. Thus, without projection, weak lensing strongly
constrains $\omde+w$ but not either parameter separately. Projection
effects break this degeneracy, because (bottom panel of
Figure~\Rf{lensing}) projection is so much more sensitive to $\omde$
than to $w$.

\Sfig{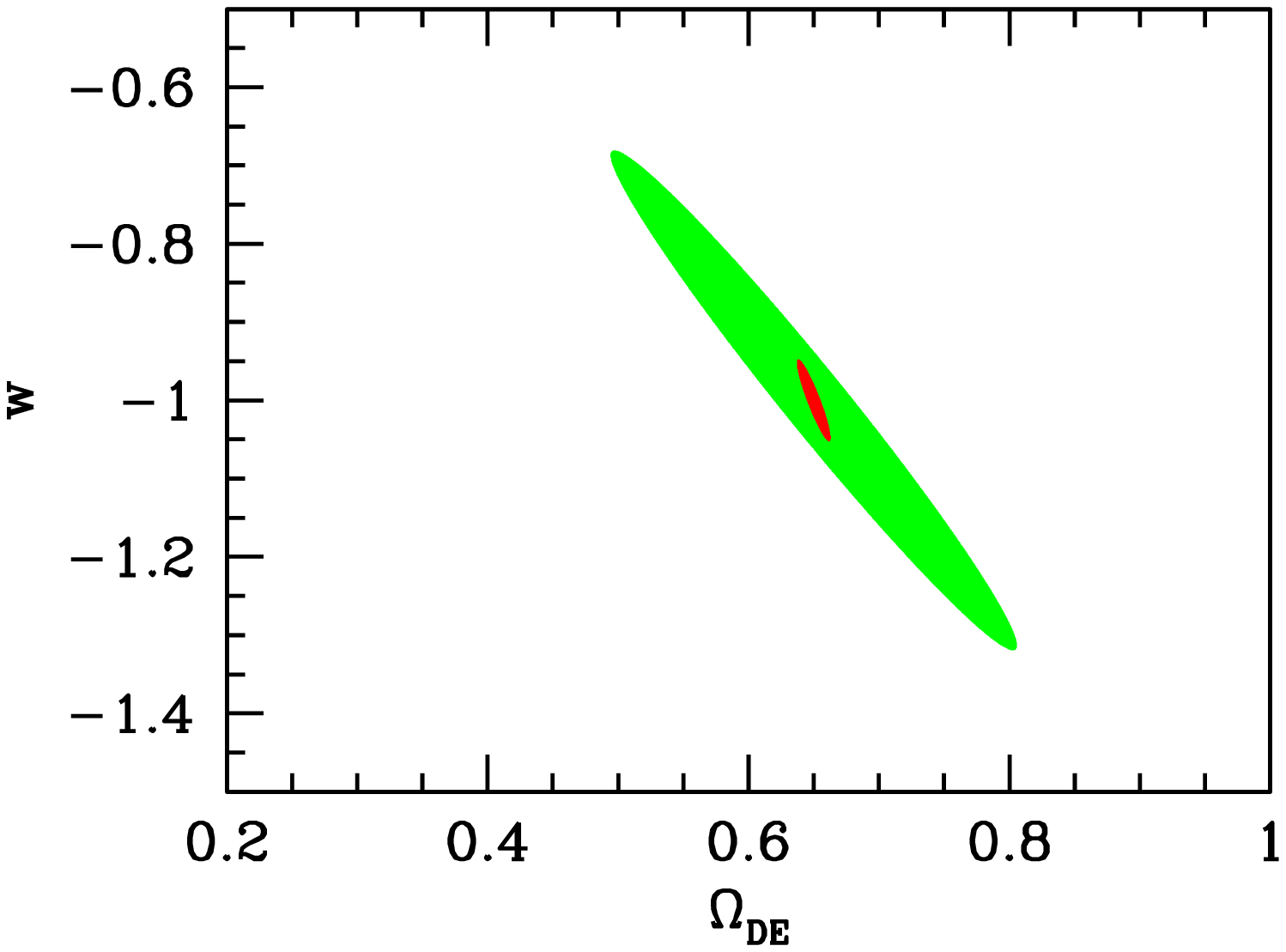}{Errors from a $4000$ deg$^2$ survey with one hundred
galaxies per square arcminute with (inner region) and without (outer
region) projection effects. Here the neutrino mass has been fixed. The
outer region uses only information from the growth function; this
information is much less constraining than including the combination
of projection.}

How do the constraints change when the uncertainty in the neutrino
mass is included?  Projection is completely independent of neutrino
mass, so the constraint on $\omde$ remains unchanged. But the limits
on the dark energy equation of state are compromised.

\Sfig{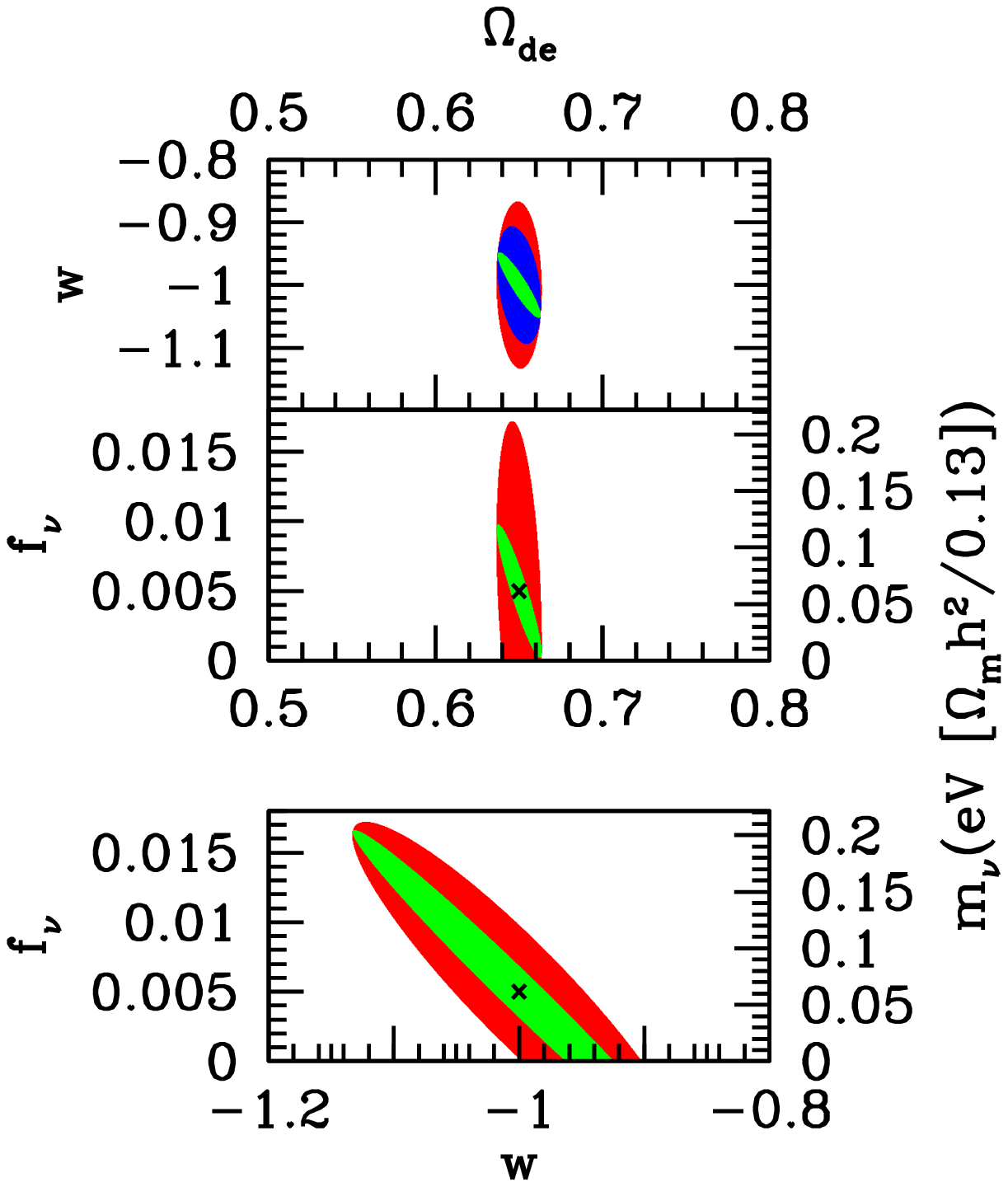}{Projected errors (all $1$-$\sigma$) on the dark
energy density, equation of state, and neutrino mass from $4000$
square degree weak lensing survey.  In each case, the innermost region
is the constraint arising from fixing the parameter not shown (e.g.,
neutrino mass in the upper panel), while the outermost constraint
comes from marginalizing over the third parameter.  The middle region
in the top panel emerges if the laboratory constraint on the neutrino
mass is $0.1$ eV.}

The top panel of Figure~\Rf{composite} indicates that the effect of
the neutrino mass is very correlated with the dark energy equation of
state. When we lack knowledge about the neutrino mass, we therefore
lose discriminatory power over the equation of state. How does the
marginalized region tighten up as external constraints on the neutrino
mass come in? If the absolute neutrino mass or its uncertainty were
constrained by a laboratory experiment (tritium endpoint~\cite{katrin}
or neutrinoless double-beta decay~\cite{doublebeta}) to be less than
$0.1$ eV, then the resulting constraint also appears in the top panel
of Figure~\Rf{composite}.  Future measurements or limits on neutrino
mass will then contribute to precision measurements of the
cosmological dark energy density and evolution.

Also of interest are the constraints on the neutrino mass. These we indicate in
the bottom panels of Figure~\Rf{composite}.  There must be a neutrino mass as
least as large as the square root of the atmospheric $\delta m^2$, i.e., $0.05$
eV. This corresponds to $f_\nu=4.1\times 10^{-3}$ for the current best value of
$\Omega_m h^2$.  Figure~\Rf{composite} shows that this limit is barely within
reach of a comprehensive weak lensing survey if we can assume that the dark
energy is a cosmological constant ($w=-1$) and that $\omde$ will be determined
by other means to within a few percent. On the other hand, if we allow for
freedom in the dark energy equation of state, then the mass limit gets worse by
a factor of order five. It is conceivable that both $\omde$ and $w$ will be
determined through other means, e.g., type Ia supernovae, cluster abundances,
or the CMB.  If so, then the neutrino mass will be detectable with weak
lensing.

These projected limits on the neutrino mass are even more powerful
than they appear.  If there are only three light neutrinos, then the
solar neutrino and atmospheric neutrino measurements constrain two
mass squared differences, as illustrated in Figure~\Rf{mass} (adapted
from Ref.~\cite{bb}). The cosmological limit discussed here is on the
sum of all neutrino masses. More stringent limits on the neutrino mass
fraction must incorporate the texture of neutrino masses, i.e., if the
neutrino masses are a standard or inverted hierarchy.  If we assume
that the probability distribution for $f_\nu$ is Gaussian with mean
$f_\nu=0.005$ and rms error corresponding to the marginalized case
$\Delta f_\nu=0.008$, then the $95\%$ confidence limit on $f_\nu$
would be $0.02$, corresponding to an upper limit on the sum of the
masses of $0.25$ eV.  Figure~\Rf{mass} shows that the region in which
all three neutrino masses are degenerate can be ruled out then by this
cosmological measurement, again even accounting for the uncertainty in
the dark energy sector.

\Sfig{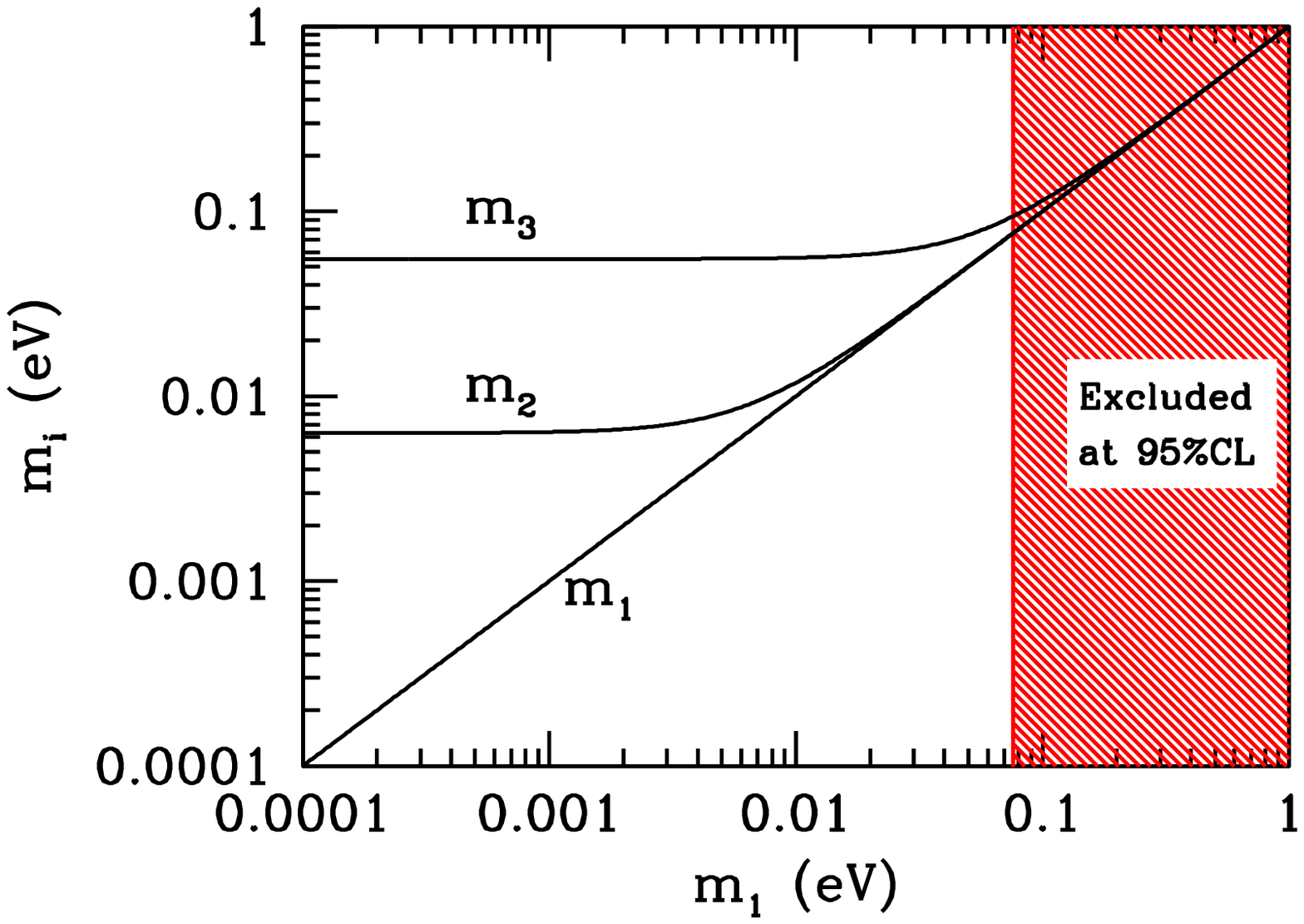}{Masses of the three neutrino species as a function of the
lightest neutrino mass in a hierarchical mass scheme~\cite{bb}. At
large $m_1$, all three must be nearly degenerate to account for the
solar and atmospheric oscillations.  The projected cosmological upper
limit excludes the shaded region at $95\%$ CL.}

We have assumed here that other cosmological parameters -- the amplitude
and slope of the primordial spectrum, $\Omega_mh^2$, and $\Omega_bh^2$ --
are fixed. If future observations constrain these parameters tightly enough
that the resulting uncertainty in the power spectrum is small, then this
assumption is valid. Figure 1 indicates that the effects we have considered here
induce of order ten percent changes in the power spectrum (which scales as the growth
function squared). Uncertainities in the power spectrum from ambiguity in the
other parameters are projected to be smaller than one percent after 
Planck\,\cite{Hu2002}.  Also, we have
not included any angular information, which would further pin down the
power spectrum of mass extremely well.  Overall, the information of
the growth of structure from radial tomography of large scale
structure via weak lensing will be a powerful method for discovering
the nature of both the dark energy and neutrino components of the
universe.

We thank John Beacom, Wayne Hu, Dragan Huterer and Eric Linder for
useful conversations.  This work is supported by the DOE, by NASA
grant NAG5-10842, and by NSF Grant PHY-0079251. \vskip -0.5cm

\newcommand\saj[3]{{Astron.\ J.\ } {\bf #1}, #2 (#3)}
\newcommand\sastronastro[3]{{Astron.\ and Astrophys.\ } {\bf #1}, #2 (#3)}
\newcommand\spr[3]{{Phys.\  Rept.\ } {\bf #1}, #2 (#3)}
\newcommand\sapj[3]{ {Astrophys.\ J.\ } {\bf #1}, #2 (#3)}
\newcommand\sprd[3]{ {Phys.\ Rev.\ } D {\bf #1}, #2 (#3)}
\newcommand\sprl[3]{ {Phys.\ Rev.\ Lett.\ } {\bf #1}, #2 (#3)}
\newcommand\np[3]{ {Nucl.\ Phys.\ B} {\bf #1}, #2 (#3)}
\newcommand\smnras[3]{{Mon.\ Not.\ Roy.\ 
	Astron.\ Soc.\ } {\bf #1}, #2 (#3)}
\newcommand\splb[3]{{Phys.\ Lett.\ } {\bf B#1}, #2 (#3)}
\newcommand\astroph[1]{{\tt astro-ph/#1}}
\newcommand\hepex[1]{{\tt hep-ex/#1}}
\newcommand\nature[3]{{Nature} {\bf #1}, #2 (#3)}


\begin{thebibliography}{99}
\bibitem{depth} To test this, we chose two redshift distributions, one
with no galaxies past $z=1.2$ and another with galaxies out to $z=3$.
The projected errors were insensitive to this change as long as the total
number of galaxies was fixed.
\end{thebibliography}
\end{document}